# Spherical-angular dark field imaging and sensitive microstructural phase clustering with unsupervised machine learning

**T P McAuliffe\***, D Dye, T B Britton

*\*t.mcauliffe17@imperial.ac.uk*

Department of Materials - Imperial College London, Prince Consort Road, London, UK

## Abstract

Electron backscatter diffraction is a widely used technique for nano- to micro-scale analysis of crystal structure and orientation. Backscatter patterns produced by an alloy solid solution matrix and its ordered superlattice exhibit only extremely subtle differences, due to the inelastic scattering that precedes coherent diffraction. We show that unsupervised machine learning (with principal component analysis, non-negative matrix factorisation, and an autoencoder neural network) is well suited to fine feature extraction and superlattice/matrix classification. Remapping cluster average patterns onto the diffraction sphere lets us compare Kikuchi band profiles to dynamical simulations, confirm the superlattice stoichiometry, and facilitate virtual imaging with a spherical solid angle aperture. This pipeline now enables unparalleled mapping of exquisite crystallographic detail from a wide range of materials within the scanning electron microscope.

## 1. Introduction

Electron backscatter diffraction (EBSD) is a common method for analysis of crystal structure and orientation in engineering materials. Typically, many thousands of electron backscatter patterns (EBSPs) are produced in a single scan of an area of interest (AOI), and there is strong motivation to develop understanding by taking advantage of the wealth of information contained within each pattern. Unsupervised machine learning allows us to explore the structure of EBSD datasets and identify latent features. In this work we compare segmentations performed with principal component analysis (PCA), non-negative matrix factorisation (NMF), and an autoencoder neural network. Performing post-segmentation analysis 'on the sphere' using a spherical harmonics approach lets us compare Kikuchi band profiles for different latent features and class averages to dynamical simulations. We find that the {100} and {111} crystallographic planes exhibit greater superlattice Kikuchi diffraction contrast than {110} and {131}. There is significant difference between simulated diffraction profiles of CoNi-Co$_3$(Al,W) and Ni-Ni$_3$Al $\gamma$ - $\gamma'$ pairings, and we confirm that our V208C Co/Ni-base superalloy matches the former. These data driven techniques enable resolution and quantitative analysis of of $\gamma$ - $\gamma'$ structures in the scanning electron microscope.

Superalloys enable modern engineering systems such as the high temperature gas turbine engine. To develop these alloys further, we require new tools to routinely analyse atomic scale ordering in microstructures. This presents a challenge at the micro-scale using EBSD, as existing analysis methods do not enable the separation of a parent crystal from its (ordered) superlattice. Often, more complicated and expensive TEM analysis using dark field imaging must be employed.

In a superalloy $\gamma$ - $\gamma'$ system the $\gamma$ matrix exhibits a face-centred cubic (FCC) structure, and $\gamma'$ superlattice precipitates with a primitive L1$_2$ structure form intragranularly below a solvus of usually 1000-1200°C. The symmetry of the FCC matrix is such that elastic (Bragg) diffraction is not observed as coming from crystallographic planes with mixed-odd-and-even Miller indices. This is not the case for $\gamma'$, and so-called 'superlattice spots' are observed in transmission electron microscope (TEM) diffraction. These are widely used to differentiate matrix from precipitate in dark field imaging. Furthermore, dark field imaging can be used to image dislocations, as the Burgers vector of a dislocation locally transforms the crystal, systematically altering the Bragg condition, and creating strong contrast for known foil normal / Burgers vector combinations. These methods have revolutionised materials science, especially in the characterisation of industrially relevant alloy deformation mechanisms, and have lead to vast improvements in creep and fatigue life of (for example) aerospace gas turbines [1–3]. However, transmission methods require thin foils, limiting the area that can be routinely studied. This motivates development of new approaches.

EBSD is performed in the scanning electron microscope (SEM), on the surface of well-polished 'bulk' materials. The method involves serial capture of 2D wide angle diffraction patterns, which contain rich structural information as they are produced from an incident electron beam scattering, diffracting, and escaping from the sample. A map can be formed when the beam is scanned in a controlled manner across the surface of the sample, and large areas at a range of step sizes can be interrogated easily.

In the superalloys, existing EBSD analysis methods index the ɣ′ and ɣ as the same (usually ɣ) phase, as unfortunately Kikuchi diffraction and the formation of EBSPs derives from extensive inelastic electron diffraction prior to elastic scattering and the formation of Kossel cones [4–6]. This means that the differences in electron scattering behaviour of ɣ′ and ɣ is extremely subtle and difficult to detect [7,8]. Diffraction patterns for the two phases, calculated using dynamical diffraction theory, are presented in Figure 1. Global EBSP simulations of ɣ and ɣ′ are performed to highlight their similarity, as well as a typical band image quality (IQ) image. The IQ map corresponds to average peak height

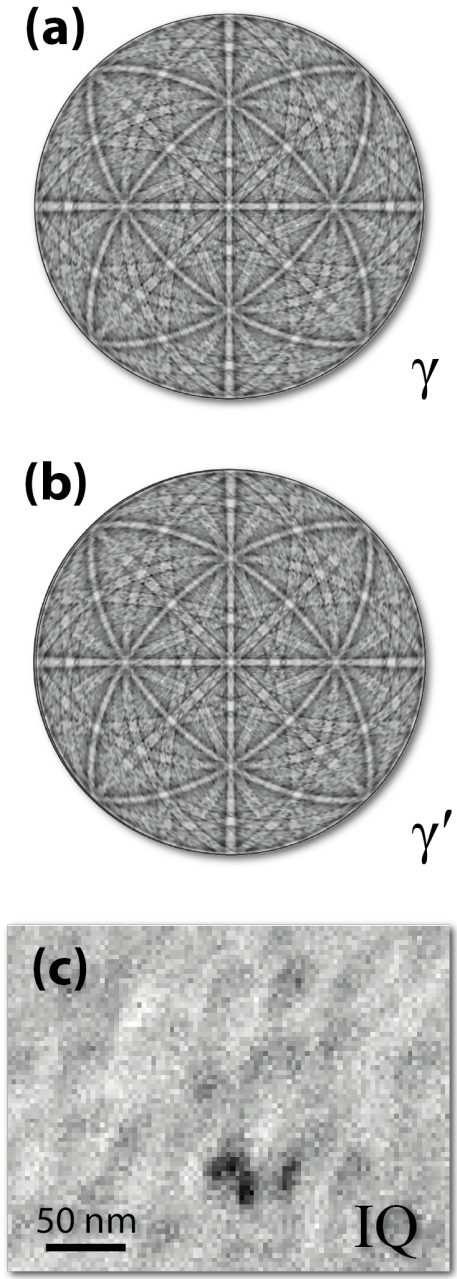

**(a)**

ɣ

**(b)**

ɣ′

**(c)**

50 nm

IQ

**Figure 1**: Similarity of EBSPs from diffracting ɣ and ɣ′. (a) and (b) show the upper halves of simulated diffraction spheres, and (c) presents a typical pattern quality map from an EBSD scan (Hough-indexed with Bruker eSprit 2.1).

of the Hough transformed EBSP, and is often used as a pseudo-backscatter mode in conventional EBSD-based imaging.

In this work, we use new, and now accessible, machine learning (ML) based-approaches to tackle this challenging segmentation problem in a Co/Ni-base superalloy. For mathematical details see Section 2.3.

PCA is a powerful tool in identifying (scan) points of self-similarity and can be used to reduce EBSD data in an observation-by-variable matrix (hence 'data matrix') down to a handful of high quality patterns. These can then be efficiently indexed [9,10]. Using PCA, we have previously combined structural information from EBSD with chemical fingerprints from simultaneous energy-dispersive X-ray spectroscopy (EDS) to characterise the phase of a microstructural constituent and separate carbide types [11,12].

NMF has seen extensive use in analysis of maps made up of (usually 1D) X-ray spectra, as the strictly positive nature of signal 'hit' counting provides an excellent boundary condition for decomposing a data matrix into its latent factors [13–15].

Autoencoder neural networks have not seen significant application in electron microscopy, and here we evaluate applicability [16–18]. They come in many flavours: deep (many-layered) or shallow, linear or convolutional. All are based on the idea of training the identity function, with information channelled through a 'bottleneck' layer containing only a few nodes. This forces the network to learn the structure of the data, with bottleneck activations representing the 'firing' of latent features. A shallow (single hidden layer) autoencoder, employed in this work, can be thought of as performing a non-linear matrix decomposition. We regularise with the L2 norm of the weights and a Kullback-Leibler divergence penalty to the hidden layer latents. Deep (convolutional) neural networks are starting to see deployment to supervised EBSD classification problems [19], but there has been little discussion of what crystallographic features neural networks are capable of learning. This is a secondary motivation for the present study.

These data science-based approaches have been performed through clustering of the captured diffraction patterns (which are 2D images captured in direct space). To provide evidence of their success, and reveal more about the microstructure of a Co/Ni-base superalloy, we analyse clustered patterns 'on the sphere'. We use this analysis to drive development of an EBSD-focussed angular resolved segmentation, and present 'spherical-angular dark field imaging'.

## 2. Methods

### 2.1 Experimental

A sample of V208C Co/Ni-base superalloy, as developed by Knop *et al* [20,21], was fabricated by vacuum arc

## (a) PCA (SVD)

## (b) NMF

## (c) Autoencoder

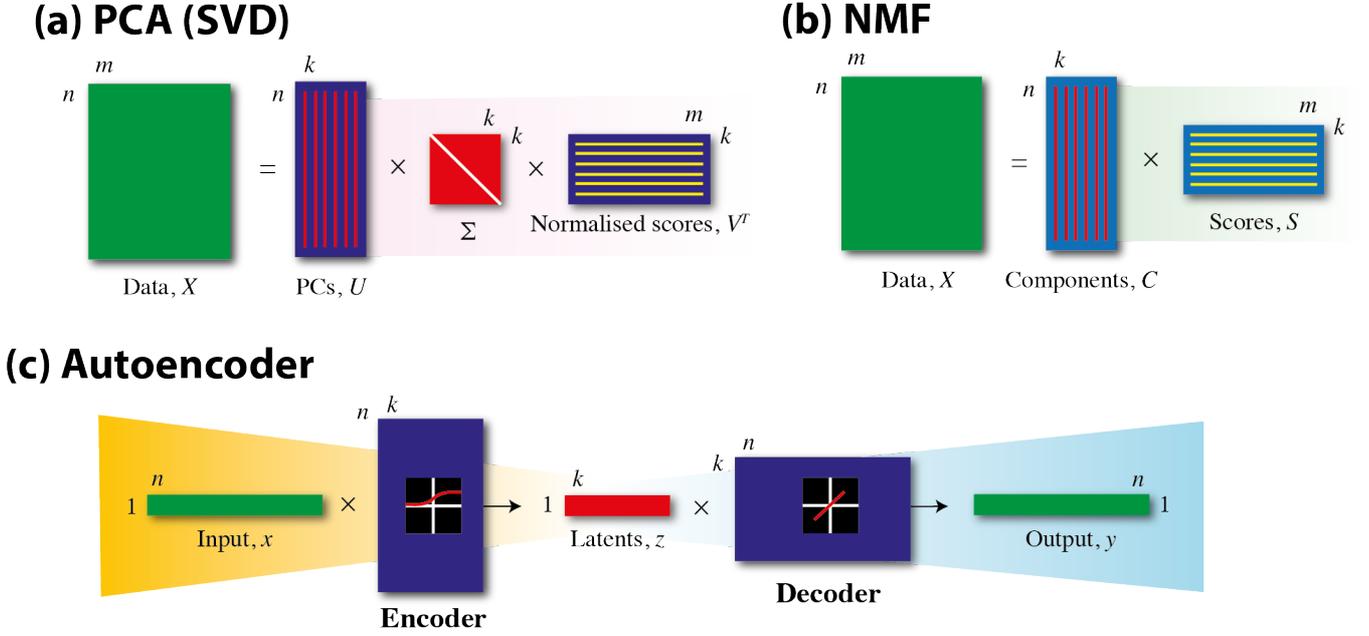

**Figure 2**: Visualisations of the matrix algebra involved in principal component analysis (a), non-negative matrix factorisation (b), and the autoencoder (c).

melting and casting, then homogenised and hot rolled at 1200°C. It was solution heat treated at 1100°C, and aged at 800°C for 4 h. The sample was prepared by standard metallographic grinding and polishing. EBSD was performed on a Zeiss Sigma FEGSEM, equipped with Bruker e⁻Flash^HD detector at 20 kV with an 120 μm aperture and high current mode, resulting in a probe current of ~10 nA. The sample was tilted to nominally 70°. For the map analysed in this work, a step size of 30 nm was used in scanning of a 2.85 μm by 2.16 μm AOI, with patterns captured at a resolution of 600-by-800 pixels. EBSPs were simulated using Bruker DynamicS [22,23], using a cut-off diffraction condition intensity of 5.0% of the maximally scattering reflector, and minimum plane spacing of 0.1 Å. These simulations were performed for CoNi (FCC solid solution), Co₃(AlW) (L1₂ ordered), Ni (FCC), and Ni₃Al (L1₂ ordered), structures. The crystal information files (CIF) for these phases are provided in the associated data bundle.

### 2.2 Data preparation

To eliminate systematic error from the analysis we normalise our EBSPs. Usually this entails subtracting the mean pixel value from each pattern and dividing by the standard deviation, which is the operation we perform for our PCA and autoencoder analysis. For NMF, as all data must be positive, we subtract the minimum value from each observation (EBSP) and divide by its standard deviation. Consequently there is an inhomogeneity in the mean value of NMF-normalised data, which is known to manifest in one of the calculated non-negative factors [15].

Without data adjustment our methods are intrinsically non-local. This means that the EBSPs could be unpacked and factorised in any order and there would be no difference to the calculated principal components (PCs), non-negative factors or autoencoder weights and biases. This has advantages, but when identifying very fine differences in latent variables it is useful to leverage localisation as additional input information to our problem. As we discuss in our previous work [11] this brings our PCA analysis closer to the NLPAR approach set forth by Brewick *et al* [24]. In the present work, we introduce a spatial weighting kernel to our pipeline that clusters patterns from local spatial neighbourhoods. This weighting kernel is applied directly to the data, and the maps can be used with our non-local algorithms as previously introduced. For our autoencoder we could have introduced a convolutional layer to the input to achieve a similar effect.

The spatial weighting kernel is described based upon the work of Guo *et al* [25], where for each EBSP $P_{ij}$ in the $i^{th}$ row and $j^{th}$ column of the AOI, the locality-corrected pattern, $P_{ij}^{local}$, is introduced as:

$$P_{ij}^{local}(r) = \sum_{k=1}^{n} \sum_{l=1}^{m} \eta \, P_{kl}$$

With kernel value:

$$\eta = \left(1 - \frac{d^2}{r}\right)^4 \ for \ d < r, else \ 0$$

Where:

$$d = \sqrt{(i-k)^2 + (j-l)^2}$$

This algorithm imposes a Gaussian-like kernel onto the AOI: patterns are averaged with weights decaying on their square separation within a kernel of consideration.

Alternative kernel functions are of course possible, be we find this one to be functional and useful. In this work we employ a kernel size, $r$, equal to three steps within the map.

### 2.3 Algorithms for learning latent factors

Graphical representations of PCA, NMF and the autoencoder are presented in Figure 2. In this section we describe each in turn.

Calculation of principal components, (Figure 2a) is equivalent to the singular value decomposition (SVD) of a ($n$-by-$m$) data matrix, $X$. This matrix has $n$ rows of variables (EBSP pixels) and $m$ columns of observations (scan points). Principal components are stored in the columns of the orthogonal matrix $U$ (the left singular vectors), and 'scores', descriptors of the extent to which a data point is represented by that vector, are given by $\Sigma V^T$ (with $V^T$ orthonormal and $\Sigma$ diagonal):

$$X = U \Sigma V^T$$

This singular value decomposition is equivalent to the eigen-decomposition of the covariance matrix, $XX^T / (n - 1)$, leading to eigenvectors in the columns of $U$ and eigenvalues $\Sigma^2$. As such, the principal components are an orthonormal set of vectors in variable space, with each minimising square distance between itself and the data in orthogonal directions. When ordered by eigenvalue (or singular value), the principal components correspond to the directions that contribute the most variance (elements of the diagonal matrix $\Sigma^2$) to the dataset, and the data matrix can efficiently approximated by a reduced number of principal component vectors ($k$):

$$X \approx X_k = \sum_{i=1}^{k} u_i \sigma_i v_i^T$$

This reduction allows us to retain the $k$ most significant features, which in an EBSD decomposition will correspond to $k$ representative Kikuchi patterns, for example one per grain, sub-grain, or precipitate. This is extremely useful when the full rank of the dataset is equal to the number of scan points (and there could be hundreds of thousands of scan points). A reduction of this type is a natural fit for problems where *a priori* we know that the AOI will only exhibit a few archetypes of Kikuchi pattern and we are not interested in intra-class variation (e.g. corresponding to crystal disorientation). The number of components to retain is often a subjective choice, but prior work has indicated that this can be selected for example using the proportion of total dataset variance contributed by the ($k$+1)[th] component [11]. In the present work, we elect to retain the components with the five largest eigenvalues.

NMF (Figure 2b) identifies a dataset decomposition with all elements strictly positive:

$$X \approx A S$$

Where $A$ and $S$ are ($n$-by-$k$) and ($k$-by-$m$) matrices, corresponding to components and scores respectively. Note that $X$ can never fully be represented by $A$ and $S$ as the rank of their product is at most $k$. This is always an approximate factorisation and inherently a less accurate one than that provided by the SVD (for the same $k$) by the Eckart-Young-Mirsky theorem [26]. $A$ and $S$ are randomly initialised, then a loss function L involving the Frobenius norm is used to alternately minimise $A$ and $S$:

$$L = \frac{1}{2} \| X - A S \|_F$$

The solution is dependent on the number of components one asks the algorithm to return. To provide comparison to our PCA pipeline, we select $k$ equal to five. Components in NMF are not orthogonal, and a unique solution is not guaranteed. Moreover, there is not an obvious geometrical analogue.

The (fully-connected) autoencoder neural network (Figure 2) we employ has three layers: an input (of dimension $n$), a single hidden layer (of dimension $k$) and the output, with the same shape as the first. A logistic activation function is applied for the sole hidden layer, with the network's outputs left unscaled after the second set of weight multiplication and bias addition.

Because the output has the same shape as the input, the network's loss function can be calculated point-wise as the mean-squared-error, combined with some regularisation terms to prevent inflation of weights and encourage sparsity. We include L2-weight regularisation (a penalty to the magnitude of the weights and biases) and a Kullback-Leibler (KL) divergence (which penalises deviation of a set of activations from a chosen distribution). Including KL divergence, applied to the hidden layer activations, encourages the network to learn a sparse representation, whereby there is a loss penalty if a training set produces too inhomogeneous a probability distribution of hidden neuron activations. Autoencoders employing KL-divergence to control the statistics of the hidden layer activations are often referred to as 'variational'. The full loss function for our autoencoder becomes:

$$L = \sum_{i=1}^{h} \sum_{j=1}^{h} (x_{i,j} - y_{ij})^2 + \lambda W + \omega K$$

Where $x$ is the function input (the unravelled EBSP, a column of $X$), $y$ the output, $W$ is the L2-regularisation and K is the Kullback-Leibler divergence term. These are modulated by hyperparameters $\lambda$ and $\omega$. W is given by:

$$W = \frac{1}{2} \left( \sum_{c=1}^{2} w_c^T w_c \right)^2$$

Where $w_c$ is the weight matrix for the c[th] layer of the network. The KL-divergence is as:

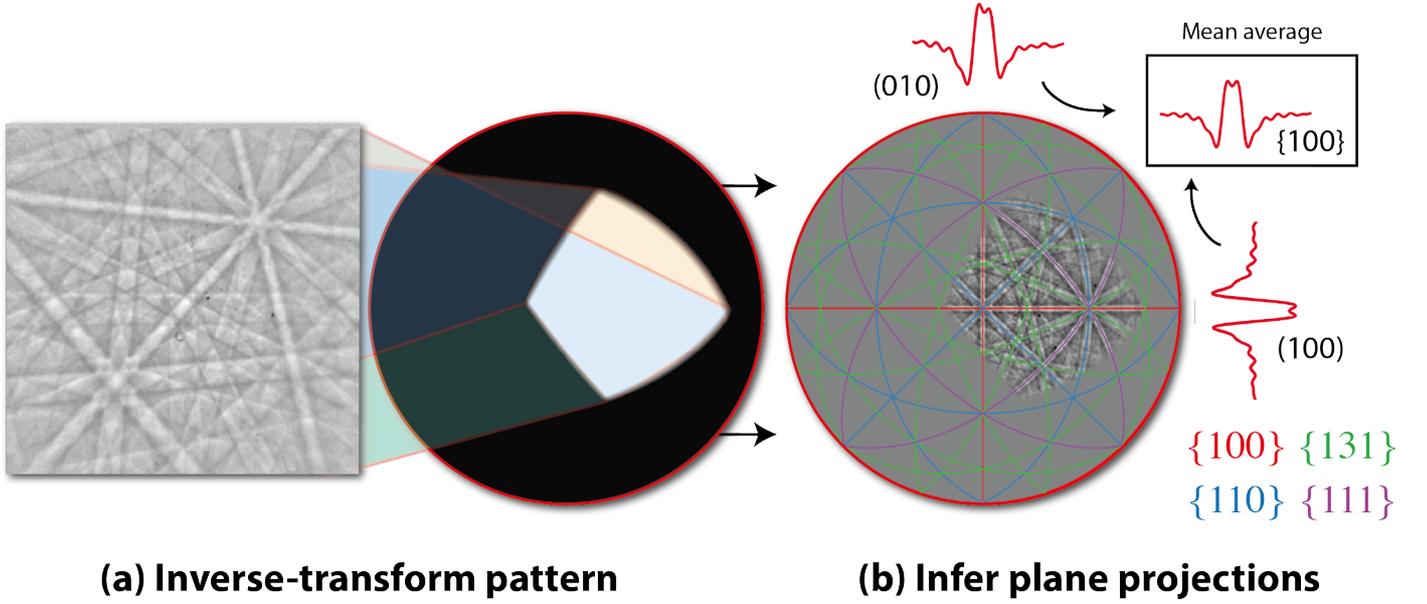

**(a) Inverse-transform pattern**

**(b) Infer plane projections**

**Figure 3**: EBSPs are re-projected onto the diffraction sphere using precise knowledge of the pattern centre and crystal orientation.

$$K = \sum_{i=1}^{k} \rho_0 [\ln \rho_0 - \ln \bar{\rho}_i] + (1 - \rho_0)[\ln(1 - \rho_0)$$
$$- \ln(1 - \bar{\rho}_i)]$$

Where $\rho_0$ is the desired sparsity proportion and $\bar{\rho}_i$ is the average likelihood of observing activating the $i^{th}$ hidden neuron:

$$\bar{\rho}_i = p(z_i \mid x) = \frac{1}{m} \sum_{s=1}^{m} z_{i,s}$$

For our $m$ observations (scan points) and hidden activations $z$. In our autoencoder we select $\rho_0$ as 0.05, with $\lambda$ and $\omega$ as 0.15. These parameters were chosen *ad-hoc* based on several training iterations. A cross-validation approach to hyperparameter selection could be employed for optimal network training, but was not deemed necessary for this work. We randomly initialise then train the network with scaled conjugate gradient descent [27] for 300 epochs (full cycles through every dataset EBSP), with network parameters updated after parsing each training data point. Scores were calculated as the values of $z$ for each example. Latent patterns are taken as the columns of the encoder matrix.

### 2.4 Spherical EBSP analysis

Analysis of the profiles of Kikuchi bands in EBSPs is inherently better suited to a spherical co-ordinate system than the gnomonic projection. In the spherical projection, the band centre is a great circle which is the plane perpendicular to the plane diffracting normal. The band can be sampled by examining subsequent small circles, each perpendicular to the diffracting plane normal. Each band profile is calculated through integration 'on-the-sphere', where Kikuchi bands have parallel edges, as opposed to integration along hyperbolic lines in the gnomonic projection [28,29].

We adopt the formulation of Hielscher *et al* [28], and re-project our measured EBSPs onto a calibrated sphere. In order to do this, we require precise knowledge of the crystal orientation at the scan point, coupled with a a precise measurement of the pattern centre. To simultaneously achieve this, we implement a simple gradient ascent algorithm, using the peak height of the cross correlation function (XCF) of a candidate EBSP and the orientation refined, simulated template from an evenly SO(3) sampled library. This 'refined template matching' approach provides a precise orientation [30]. Starting with an initial estimate of the pattern centre and crystal Euler angles (from Bruker Esprit 2.1), we simulate templates with increments in PCX, PCY and detector distance (DD), following conventions of Britton *et al* [31]. We infer the gradient in XCF peak height with respect to PCX, PCY and DD to generate an updated centre, and template match to get an updated orientation. This procedure is implemented in MATLAB and iterated with decreasing step size in PCX, PCY and DD to generate a highly accurate pattern centre and orientation.

We use this geometry and crystal orientation to re-project candidate EBSPs onto the sphere for subsequent analysis. This is achieved by calculating a function $f$ with respect to diffraction directions $\boldsymbol{\xi}$ that maps the inverse gnomonic projection. This follows the expansion:

$$f(\boldsymbol{\xi}) = \sum_{a=0}^{N} \sum_{b=-\beta}^{\beta} \hat{f}(a, b) \, Y_a^b(\boldsymbol{\xi})$$

Where $Y_a^b$ are the spherical harmonic functions, and $\hat{f}(a, b)$ are the Fourier coefficients of $f$. $N$ corresponds to the degree of harmonic employed for the expansion. Several approaches for calculation of the Fourier coefficients have been discussed by Hielscher *et al* [28], and in this work we employ the 'quadrature' method with $N = 256$.

| Factor | 1 | 2 | 3 | 4 | 5 |
|--------|------|------|------|------|------|
| 1 | **0.2036** | 0.0595 | 0.0392 | 0.0630 | 0.0064 |
| 2 | 0.0595 | **0.1930** | 0.0509 | -0.0179 | 0.0301 |
| 3 | 0.0392 | 0.0509 | **0.1917** | 0.0205 | 0.0605 |
| 4 | 0.0630 | -0.0179 | 0.0205 | **0.2010** | 0.0945 |
| 5 | 0.0064 | 0.0301 | 0.0605 | 0.0945 | **0.2043** |

**Table 1**: Covariances of autoencoder latent factor scores. Shown in **bold** are the intra-factor variances.

The diffraction pattern, now on the sphere, can be analysed. The band profiles can be extracted through projections of relevant crystallographic planes using the *MTEX* orientation analysis MATLAB package. This pipeline is presented in Figure 3.

With this calculated projection and the Kikuchi band profiles, we can integrate intensity in spherical coordinates. These correspond to summing the path-normalised sums of the small circles of the diffraction sphere, varied along the opening angle around a plane projection. Specifically, this is an integration with respect to all the possible rotations around the plane normal, $R$ about angles $\theta$ [28]:

$$\phi(\xi) = \int_0^{2\pi} \Psi(\, R(\theta)\, \xi\,) d\theta$$

With $\phi(\xi)$ the resultant band profile and $\Psi$ the spherically projected pattern. We perform this analysis for both global simulations (in which we calculate the full $\Psi$) and re-projected patterns in experimental co-ordinates, for which $\Psi = f(P)$, with $P$ a single (square) pattern. In this work, if multiple {hkl} plane projections are present in the field of view we mean-average the profiles.

### 2.5 Computation, software, and data

The analysis presented in this work was performed on a 64-bit Windows 2019 Server PC, with an Intel ® Xeon ® Gold 6138 CPU and 256 GB of RAM. The *SciKit-Learn* Python 3.7 package was used for PCA and NMF decompositions, and the autoencoder developed in MATLAB 2019b with the *Statistics & Machine Learning* toolbox. NMF and PCA EBSP dataset decomposition routines have been implemented in the *ebspy* Python package. Our pattern centre refinement algorithm is available in *AstroEBSD*, written in MATLAB. These repositories are available open access and can be found at  and  respectively. Band analysis workflows and spherical-angular dark field imaging will be included in *AstroEBSD* upon article acceptance.

## 3. Results & Discussion

In a Co/Ni-base superalloy sample known to contain a high γ′ volume fraction, we collect EBSPs across an area of interest of 2.85-by-2.16 μm, employing a scanning step size of 30 nm. Details of data preparation, algorithm hyperparameter choice and regularisation are included in Section 2.

### 2.1 Decomposition evaluation

We compare the latent factors uncovered by unsupervised learning of our AOI, evaluating coefficients (latent pseudo-EBSPs) and corresponding scores uncovered by PCA, NMF and our autoencoder neural network with hyperparameters discussed in Section 4.3. All three approaches are able to extract subtle, physically significant features.

The first two principal components (PCs) (Figure 4 a-1,2; b-1,2) represent information specific to γ and γ′. PCs 1 and 2 provide reasonable distinction between precipitate and matrix, and we attribute contrast in the higher order components to sample topography (and subsequent impact on electron exit angle). This appears to be the case especially for PC 5, where there is a horizontal spatial mode. Such influence is not insignificant, as reflected in the magnitude of the scores of the higher order (topography-driven) components being of a similar magnitude to the structure-derived differences for PCs 1 and 2.

The solution identified by NMF appears physical in origin, as compared to the statistical solution from PCA. These factors are not ordered by dataset contribution (as the PCs are). Human based analysis of the NMF solutions reveals high contrast segmentation of γ and γ′ in Factor 3 (Figure 4 c-3; d-3), with signal from the γ regions positively aligning to a vector with γ′ to γ clusters in variable space. We attribute Factor 1 to variation in background (as discussed in Section 2.2 our positivity condition necessitates a roaming EBSP mean), and the remaining factors to a combination of weak variations in

# Principal Component Analysis

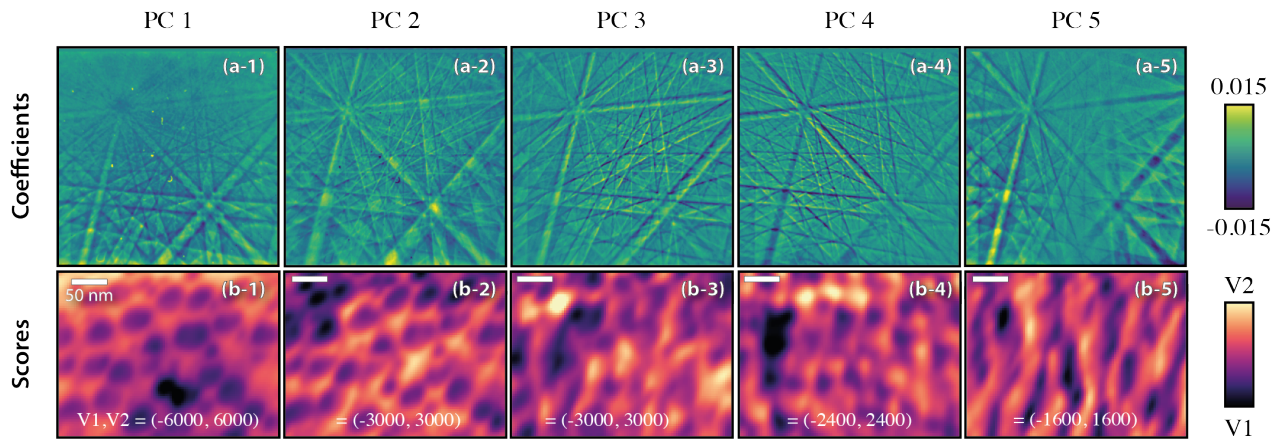

# Non-negative Matrix Factorisation

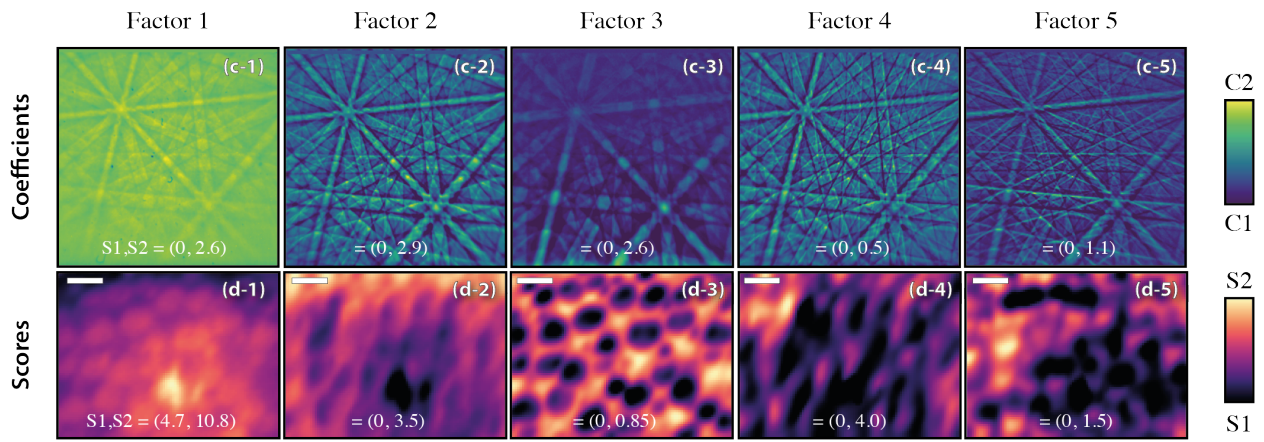

# Autoencoder neural network

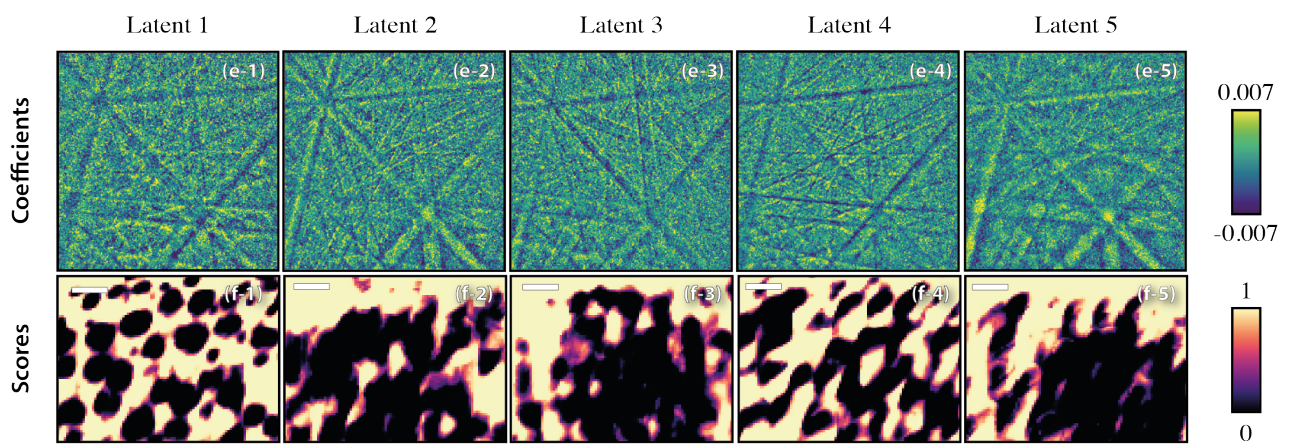

# Reference

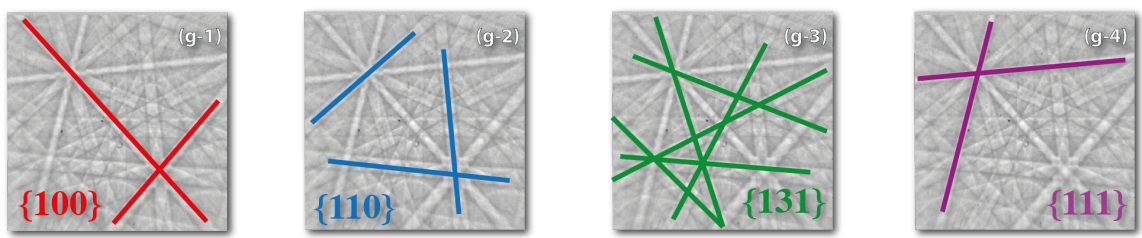

**Figure 4**: Latent factors identified by our three unsupervised machine learning algorithms. PCA - (a,b), NMF – (c,d), and the autoencoder (e,f). Kikuchi bands are labelled in (g) to show corresponding plane projections.

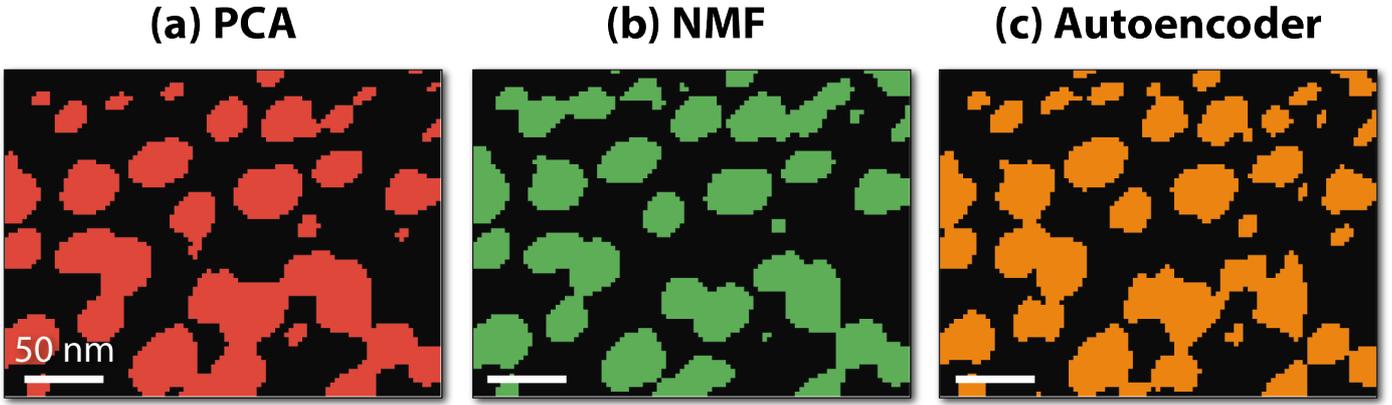

**Figure 5**: Clustering performed on the PCA (a), NMF (b), and autoencoder (c) decompositions, in order to identify ɣ (black) and ɣ′ (coloured).

topological mode as in PCA. The factor aligned with precipitate difference (Factor 3) concentrates intensity in the band interiors, which is where the band contrast lies as we observe in Figure 4. Finally, the decomposition identified by the autoencoder neural network appears as heavily binarised versions of those seen in the matrix decompositions. Human based analysis of the autoencoder solution can be used to attribute Latent 1 to our expected deviation between precipitate and matrix, with the others not being so easily interpretable.

Binarisation in Latent 1 is desirable, as we are attempting to separate two very similar crystal structures. For further insight into the network output we present in Table 1 the covariances of the latent factor scores. As requested the variance in each factor (the diagonal components) are approximately 0.2, with the off-diagonal components smaller. There remains a fairly substantial degree of covariance (for example Factors 4 and 5), even after significant KL-divergence regularisation. As most notably comparable with NMF (PCA does not present a single latent factor solely attributable to ɣ′) contrast between matrix and precipitate is mostly contained to the band interiors, observed in Latent 1(Figure 4 e-1; f-1). The sense of this vector is opposite to that in the NMF, despite similar sense in ɣ / ɣ′ separation in scores. We expect this is due to the difference in normalisation required by NMF and which side of the mean the vector is operating from. The latent signals are much noisier than those identified with the linear methods, despite a low ultimate error rate and very accurate reconstruction.

*2.2 Superlattice segmentation*

All three approaches are able to identify subtle differences between ɣ and ɣ′. Firstly, analysing the PCA-reduced dataset, we binarise at the 0.38 quantile point (subjectively identified) on the sum of the first three PCs (which we interpret to contain significant matrix / precipitate contrast). For the NMF and autoencoder separations, we binarise at the 0.38 quantile on Factor 3 and Latent 1 respectively. The segmentations are presented in Figure 5. All three approaches lead to reasonable and spatially consistent classification. Subsequently, we take average measured EBSPs from each of the classified regions and compare to dynamical simulations 'on-the-sphere' after Hielscher *et al* [28] (for details see Section 2.4) in order to correct for hyperbolic divergence of as-measured (gnomonically projected) Kikuchi patterns. The spherically integrated profiles, $\phi_{ij}^{hkl}$, of crystal plane families {hkl} for each of the ɣ - ɣ′ class-average EBSPs are plotted in Figure 6. These are directly compared with dynamical simulations (for a replica of the crystal orientation and camera geometry). Due to the similarly good performance in classification, the cluster-average patterns are extremely similar for PCA, NMF and autoencoder approaches. For brevity we only present the PCA cluster averages, and include the rest as Supplementary Figure 1.

Figure 6a shows the same pattern in ɣ and ɣ′ intensities for the clustered experimental EBSPs as for the dynamical simulations of the CoNi-Co₃(AlW) system: ɣ′ generally diffracts less at small opening angles than ɣ, as was qualitatively observed in the latent factor backscatter patterns (Figure 4). This is the case across the AOI for the {100}, {110}, {131}, and {111} band profiles. The differences between matrix and precipitate profile, presented in Figure 6b, shows the experimental (clustered) EBSPs retain consistently the same sign in profile difference as the Co pairing, and are opposite in sign to the Ni pairing. In order to confirm that this observation is not an artefact of our detector undersampling the diffraction sphere, we perform global simulations in addition to the inverse-gnomonically projected single templates analysed in Figure 6. These simulations show the same fingerprint, and are included as Supplementary Figure 2. The relative differences in contrast between superlattice and matrix are discussed further in Section 2.3.

We attribute the difference in scattering behaviour between Ni and Co-base systems to chemical segregation between matrix and precipitate. In Co-base superalloys the ɣ′ precipitates require stabilisation with W to prevent formation of extraneous microstructural phases such as B2 CoAl and DO₁₉. This necessitates segregation of W to the precipitates, and will result in a greater tendency for inelastic electron scattering [8,32,33]. This in turn reduces the elastically diffracted signal for the ɣ′ relative to the γ. The lack of heavy element segregation to γ′ in the Ni / Ni$_3$Al system leads to a reversal in contrast, which is not observed the experimental patterns. Winkelmann & Vos [34] have previously established the importance of the atomic species and degree of localisation of the scattering inelastic source in Kikuchi band formation. Our results, and attribution of the segregation of W to our γ′ system, agree with their conclusion that eventual measured intensity distributions are sensitive to scattering of the incoherent point source in the unit cell.

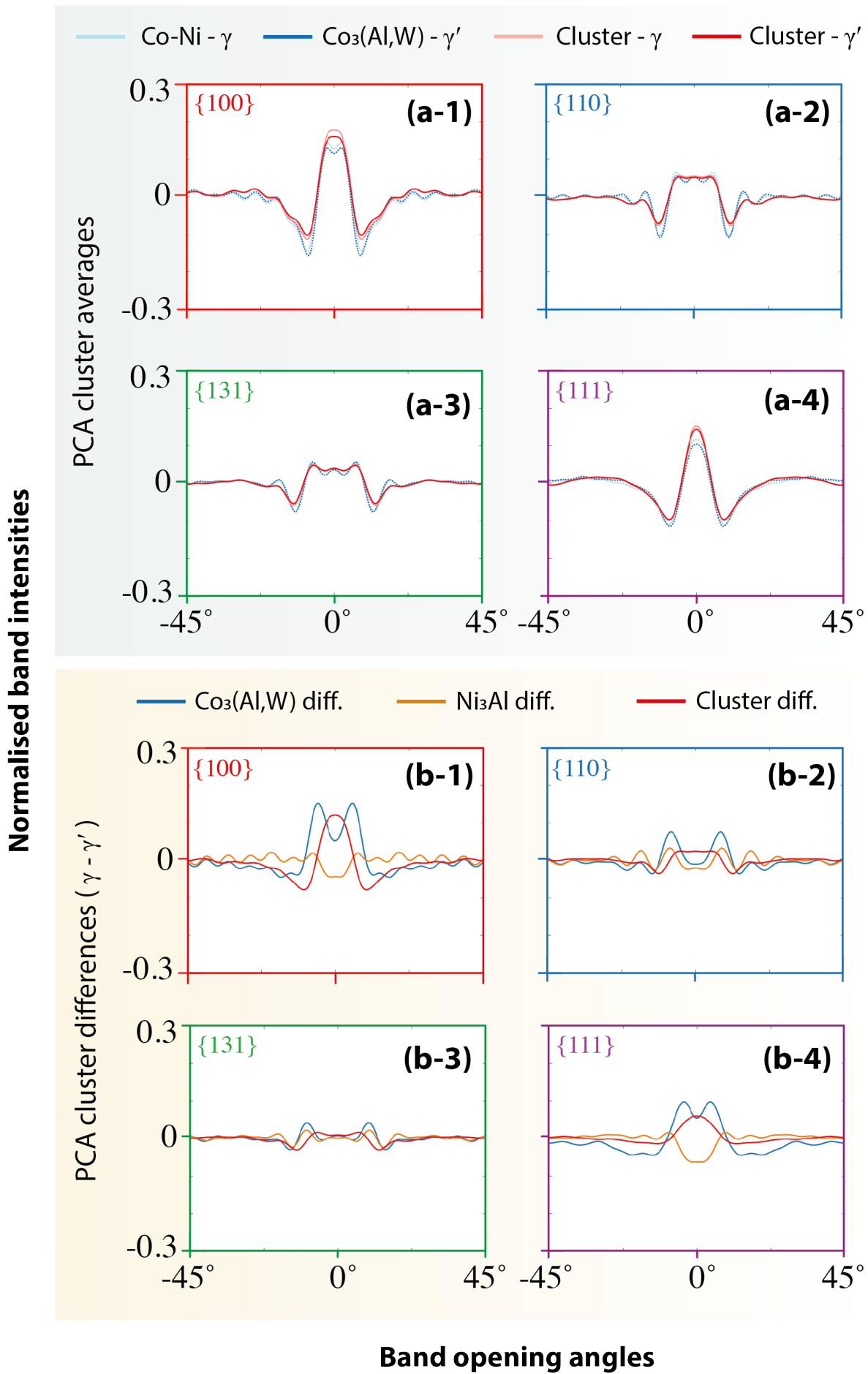

**Figure 6**: Comparison between Kikuchi band profiles obtained from (a) averages of the PCA-identified clusters, and (b) differences between profiles for simulated CoNi-Co₃(Al,W) and Ni-Ni₃Al, and PCA-cluster ɣ - ɣ' systems.

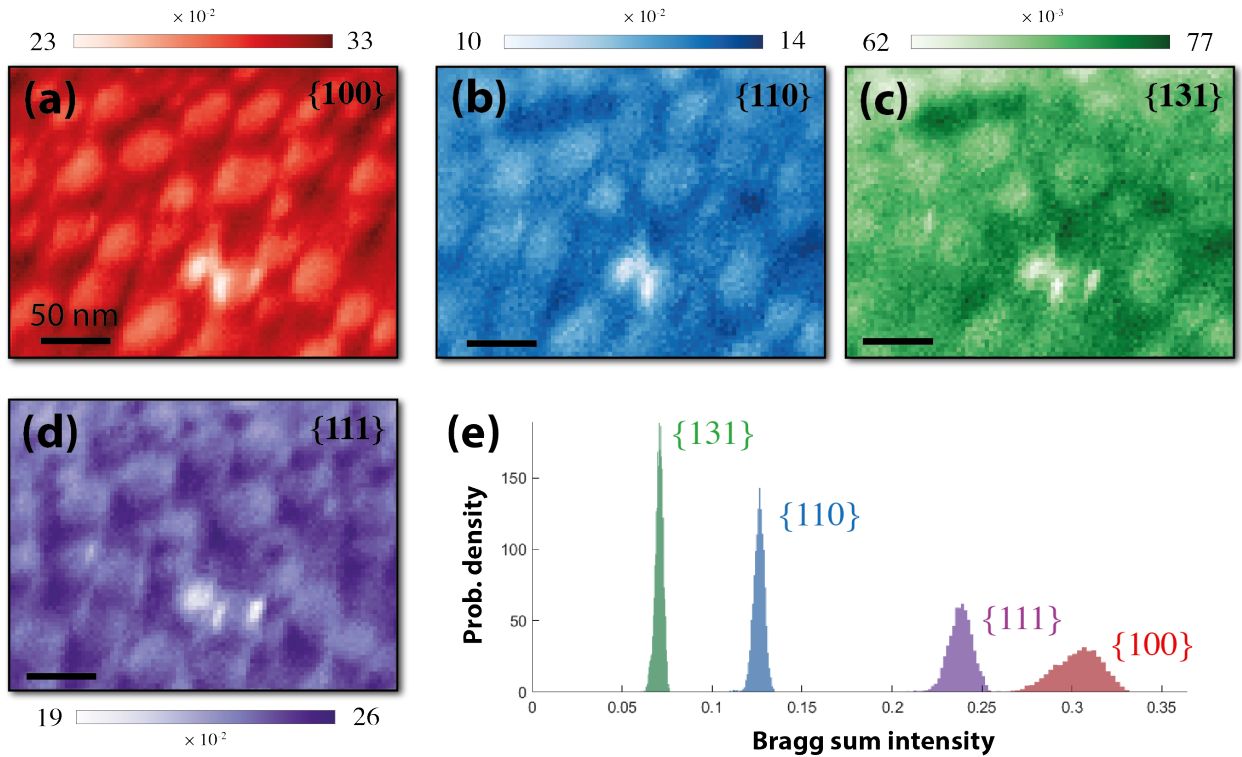

**Figure 7**: Spherical-angular dark field imaging (Bragg summations) of the dataset: (a-d) integrated intensities of the corresponding band profiles , and (e) probability density distribution (histogram normalised by number of observations and bin width) of the calculated intensities for each of the diffraction conditions.

### 2.3 Spherical-angular dark field imaging

In order to directly observe the γ - γ′ intensity polarisation seen in the cluster-average EBSPs, we calculate the profile sums within one Bragg angle of the plane projection (integrating +/- the Bragg angle from 0° in the profile scheme of Figure 6), for every scan point and plane family. Specifically,

$$S_{ij}^{hkl} = \int_{\theta=0-\theta_{bragg}}^{\theta=0+\theta_{bragg}} \phi_{ij}^{hkl} \, d\theta$$

This is evaluated at scan point $i,j$, with spherically projected profile $\phi_{ij}^{hkl}$, plane family {hkl}, and band opening angle $\theta$. This enables us to generate virtual 2D microstructural images based from specific diffraction conditions, similarly to 'virtual dark field' analyses commonly performed in the TEM community [35–37]. This is a more informed approach than previous dark field EBSD-based methods [38,39]. Analysis 'on-the-sphere', where we account for hyperbolic divergence of the

Kikuchi bands in the gnomonic projection, enables windowing of specific diffraction based contrast variations. This is only possible as the averaging profiles from all integrated {hkl} Kikuchi bands are directly employed to increase signal to noise, and improving confidence in the crystallographic origin of our contrast. Contrast is verified with analysis of simulations.

Analysis of the {100}, {110}, {131}, and {111} conditions is presented in Figure 7. There is a wider spread in intensity for the {100} and {111} integrations. The magnitudes of the peaks in Figure 6 (b-1, b-4) are greater than those in (b-2, b-3), in turn agreeing with dynamical simulation. The probability density function, Figure 7e, accordingly displaying a greater spread of intensities for these high-contrast conditions than {110} and {131}. In Figure 7e there are two superposed intensity distributions (which we do not resolve here) for each diffraction condition, corresponding to signals from the matrix and precipitate. The wider spread in intensity for {100} and {111} virtual crystallographic images is a result of better separated average intensities, as we know from the cluster-average profiles in Figure 6.

### 2.4 Consistency and discussion of clustering approaches

In order to compare our clustering approaches for separation of precipitate from matrix, we calculate and apply a simple γ - γ′ normalised contrast metric, $C$, for each diffraction condition family and clustering approach:

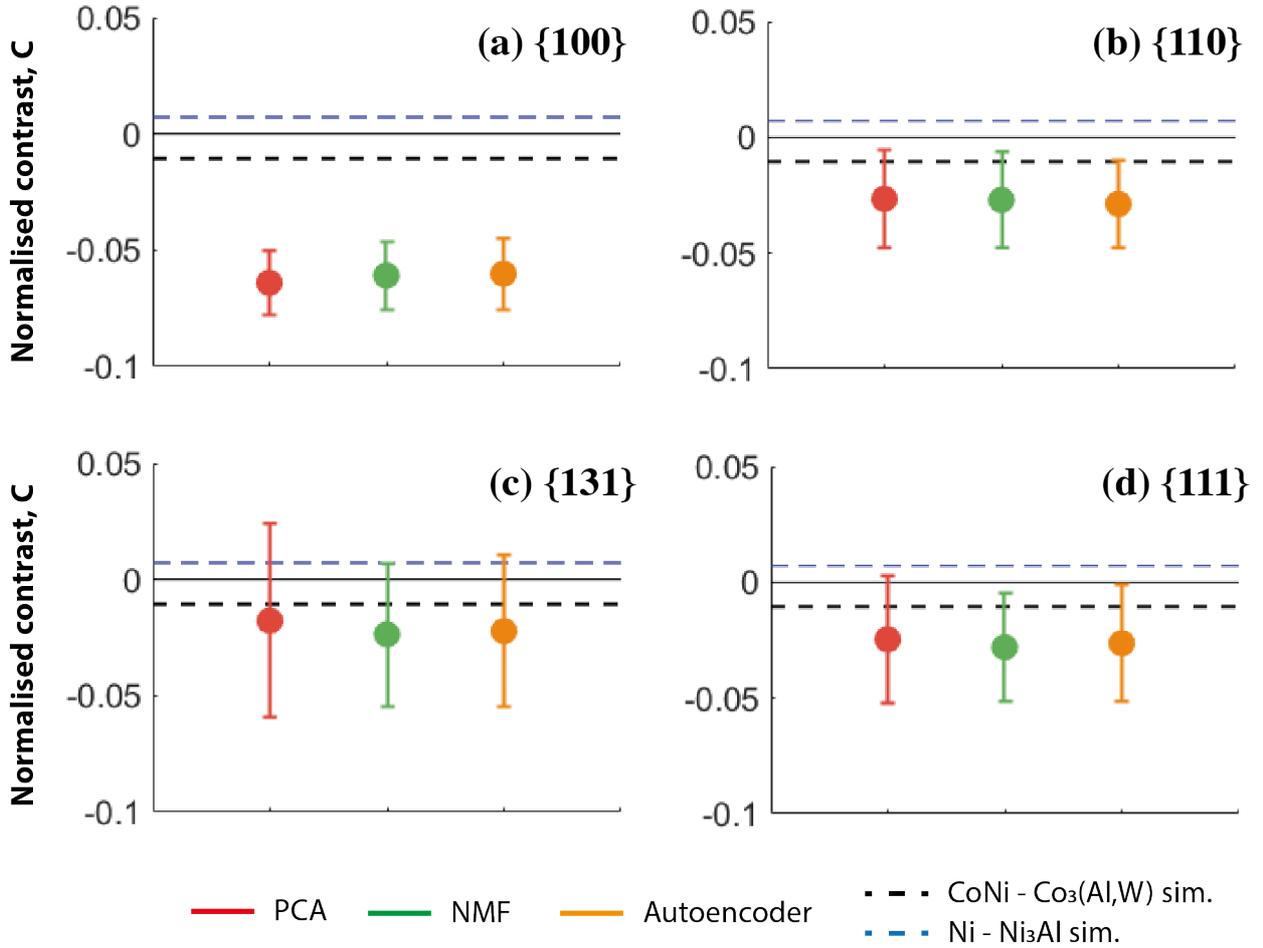

**Figure 8**: Normalised contrast metrics calculated for each of the diffraction conditions we observe, compared to CoNi-Co₃(Al,W) and Ni-Ni₃Al ɣ - ɣ′ pairing dynamical simulations.

$$C = \frac{S_{\gamma'}^{hkl} - S_{\gamma}^{hkl}}{S_{\gamma'}^{hkl} + S_{\gamma}^{hkl}}$$

For $S_{phase}^{hkl}$ in the first instance corresponding to the mean value of ɣ and ɣ′ segmented regions of $S_{ij}^{hkl}$. $C$ is calculated for each of the segmentation approaches (spatially resolved in Figure 5). We also calculate $C$ for the simulated CoNi-Co₃(Al,W) and Ni-Ni₃Al ɣ - ɣ′ pairings. These, along with errors propagated from standard error in the means of the corresponding plane-specific integrated intensities, are presented in Figure 8.

These contrast metrics show that all three dataset segmentations capture the negative contrast predicted for the CoNi-Co₃(Al,W) system. We can conclude that the ɣ′ in this alloy has crystallographic behaviour closer to the Co₃(Al,W) archetype than Ni₃Al, a characteristic that has historically required TEM to observe.

Of our approaches, NMF seems to achieve the best consistency in segmentation (resulting in smaller error-bars in Figure 8), and broadly the only approach able to present consistently negative contrast across all diffraction conditions fully within error. This is due to the algorithm's convergence to a basis that explicitly includes a factor corresponding to a ɣ - ɣ′ vector in variable space. This was not the case for PCA, which did not identify a component as well aligned to this crystallographic difference.

The autoencoder identified a similar latent factor to NMF, well aligned to a crystallographic difference vector. However, the segmentation was not as well spatially resolved. This is likely due to the activation function we employ, which due to the extreme similarity between matrix and precipitate EBSPs experienced a challenging task in where to correctly assign the domain of the logistic sigmoid function to observe the necessitated (Kullback-Leibler regularised) variation in latent score. Our autoencoder implementation with one hidden layer and no convolution filters compares favourably to PCA and NMF. With supervised structure classification using deep neural networks beginning to see application to EBSD data, we believe it is valuable to investigate specifically what can be learned by simpler architectures. After 300 epochs of (over-) training the latent representation our simple network has learned is noisy, despite very accurate reconstructions. The ability of the network to learn EBSP-specific features (such as zone axis contrast, band intensity profiles) likely further diminishes with max-pooling and convolution operations.

Spherical-angular dark field imaging presents exciting possibilities for microstructural analysis. Conventional micro-scale SEM imaging modes (usually backscatter or secondary electron) are naïve to detected electron energy and geometry. Superalloys, due to their structural and chemical similarity in total inelastic scattering propensity, therefore often require chemical etching and surface modification to generate suitable contrast for evaluation. This results in highly subjective analysis, as knowledge of how the surface is modified can be extremely complicated and vary. Direct use of scattering and diffraction data reduces this uncertainty and we have shown that collecting intensity 'on-the-sphere' generates contrast between precipitate and matrix at specific diffraction conditions. This contrast is due to intensity differences derived from chemical segregation. Averaging over ML-derived segmentations of the dataset amplifies the signal to noise for these subtle variations and provide a comparison metric between microstructural constituents. We expect that such an approach will prove useful for similarly challenging crystallographic similarity problems in EBSD, such as martensite characterisation and carbide type differentiation.

## 4. Conclusions

In this work we have investigated segmentation of γ matrix from γ′ precipitate in a Co/Ni-base superalloy, using unsupervised machine learning. EBSPs were successfully clustered using principal component analysis, non-negative matrix factorisation, and an autoencoder neural network. We draw the following conclusions:

- All three approaches are suitable for identification of very subtle differences in EBSP band contrast resulting from superlattice ordering. NMF provides the most physically justifiable basis, including a factor that explicitly aligns with region ordering. The autoencoder finds a similar feature, but with worse spatial fidelity and noisy latents. PCA finds a basis that includes reasonable superlattice contrast in the strongest principal components, with higher order terms appearing to represent sample topography.

- Segmentations from all three approaches explicitly show less intense diffraction at the band cores in superlattice ($L1_2$) γ′ than in matrix (FCC) γ. This agrees with simulations of a CoNi-$Co_3$(Al,W) system, and is opposite in sense to simulations of a Ni-$Ni_3$Al system. We attribute this behaviour to reduced elastic scattering in γ′ where heavier elements (such as W) tend to segregate.

- Virtual crystallographic imaging of the area of interest (summing intensity within one Bragg angle

of the plane projection, accounting for hyperbolic band divergence) shows greater normalised superlattice/matrix contrast for {100} and {111} diffraction conditions than {110} and {131} for all three segmentation approaches. NMF provides consistently the lowest standard error in this contrast metric for the bands we have integrated.

## 5. Acknowledgements

We are grateful to the developers of the *MTEX* orientation analysis package and Hielscher *et al* [28] for making their code available open access under an MIT license at github.com/mtex-toolbox/mtex-paper.

All the authors acknowledge support from the Rolls-Royce plc - EPSRC Strategic Partnership in Structural Metallic Systems for Gas Turbines (EP/M005607/1), and the Centre for Doctoral Training in Advanced Characterisation of Materials (EP/L015277/1) at Imperial College London. TBB additionally acknowledges the Royal Academy of Engineering for funding his research fellowship. DD acknowledges funding from the Royal Society for his industry fellowship with Rolls-Royce plc. We thank the Harvey Flower Microscopy Suite at Imperial College London for access to microscopy facilities.

## 6. References

[1] V.A. Vorontsov, L. Kovarik, M.J. Mills, C.M.F. Rae, High-resolution electron microscopy of dislocation ribbons in a CMSX-4 superalloy single crystal, Acta Mater. 60 (2012) 4866–4878. doi:10.1016/j.actamat.2012.05.014.

[2] C.M.F. Rae, N. Matan, R.C. Reed, The role of stacking fault shear in the primary creep of [001]-oriented single crystal superalloys at 750°C and 750 MPa, Mater. Sci. Eng. A. 300 (2001) 125–134. doi:10.1016/S0921-5093(00)01881-3.

[3] R.C. Reed, N. Matan, D.C. Cox, M.A. Rist, C.M.F. Rae, Creep of CMSX-4 superalloy single crystals: Effects of rafting at high temperature, Acta Metall. 47 (1999) 3367–3381.

[4] D.J. Dingley, S.I. Wright, Determination of crystal phase from an electron backscatter diffraction pattern, J. Appl. Crystallogr. 42 (2009) 234–241. doi:10.1107/s0021889809001654.

[5] G. Nolze, A. Winkelmann, Crystallometric and projective properties of Kikuchi diffraction patterns, J. Appl. Crystallogr. 50 (2017) 102–119. doi:10.1107/s1600576716017477.

[6] S. Zaefferer, A critical review of orientation microscopy in SEM and TEM, Cryst. Res. Technol. 46 (2011) 607–628. doi:10.1002/crat.201100125.

[7] A.J. Wilkinson, T. Ben Britton, Strains, planes, and EBSD in materials science, Mater. Today. 15

(2012) 366–376. doi:10.1016/S1369-7021(12)70163-3.

[8] C.B. Carter, D.B. Williams, Transmission Electron Microscopy, Second, Springer, 2009.

[9] A.J. Wilkinson, D.M. Collins, Y. Zayachuk, R. Korla, A. Vilalta-Clemente, Applications of Multivariate Statistical Methods and Simulation Libraries to Analysis of Electron Backscatter Diffraction and Transmission Kikuchi Diffraction Datasets, Ultramicroscopy. 196 (2018) 88–98. doi:10.1016/j.ultramic.2018.09.011.

[10] L.N. Brewer, P.G. Kotula, J.R. Michael, Multivariate statistical approach to electron backscattered diffraction, Ultramicroscopy. 108 (2008) 567–578. doi:10.1016/j.ultramic.2007.10.013.

[11] T.P. McAuliffe, A. Foden, C. Bilsland, D. Daskalaki-Mountanou, D. Dye, T.B. Britton, Advancing characterisation with statistics from correlative electron diffraction and X-ray spectroscopy, in the scanning electron microscope, Ultramicroscopy. (2020) 112944. doi:10.1016/j.ultramic.2020.112944.

[12] T. McAuliffe, L. Reynolds, I. Bantounas, T. Britton, D. Dye, The Use of Scanning Electron Beam-based Phase Classification as a Crucial Tool in Alloy Development for Gas Turbine Engine Applications, Microsc. Microanal. 25 (2019) 2402–2403. doi:10.1017/s1431927619012741.

[13] J. Spiegelberg, S. Muto, M. Ohtsuka, K. Pelckmans, J. Rusz, Unmixing hyperspectral data by using signal subspace sampling, Ultramicroscopy. 182 (2017) 205–211. doi:10.1016/j.ultramic.2017.07.009.

[14] M. Shiga, K. Tatsumi, S. Muto, K. Tsuda, Y. Yamamoto, T. Mori, T. Tanji, Sparse modeling of EELS and EDX spectral imaging data by nonnegative matrix factorization, Ultramicroscopy. 170 (2016) 43–59. doi:10.1016/j.ultramic.2016.08.006.

[15] M.W. Berry, M. Browne, A.N. Langville, V.P. Pauca, R.J. Plemmons, Algorithms and applications for approximate nonnegative matrix factorization, Comput. Stat. Data Anal. 52 (2007) 155–173. doi:10.1016/j.csda.2006.11.006.

[16] G.E. Hinton, R.R. Salakhutdinov, Reducing the Dimensionality of Data with Neural Networks, Science (80-. ). 313 (2006) 504–507.

[17] H. Valpola, From neural PCA to deep unsupervised learning, Adv. Indep. Compon. Anal. Learn. Mach. (2015) 143–171. doi:10.1016/b978-0-12-802806-3.00008-7.

[18] W. Wang, Y. Huang, Y. Wang, L. Wang, Generalized Autoencoder: A Neural Network Framework for Dimensionality Reduction, CVPR Work. (2014) 490–497.

[19] K. Kaufmann, C. Zhu, A.S. Rosengarten, D. Maryanovsky, T.J. Harrington, E. Marin, K.S. Vecchio, Crystal symmetry determination in electron diffraction using machine learning, Science (80-. ). 568 (2020) 564–568.

[20] D. Dye, M. Hardy, H. Yan, M. Knop, H. Stone, EP2821519B1, 2017.

[21] M. Knop, P. Mulvey, F. Ismail, A. Radecka, K.M. Rahman, T.C. Lindley, B.A. Shollock, M.C. Hardy, M.P. Moody, T.L. Martin, P.A.J. Bagot, D. Dye, A new polycrystalline Co-Ni superalloy, JOM. 66 (2014) 2495–2501. doi:10.1007/s11837-014-1175-9.

[22] A. Winkelmann, Dynamical Simulation of Electron Backscatter Diffraction Patterns, Electron Backscatter Diffr. Mater. Sci. (2009). doi:10.1007/978-0-387-88136-2 2.

[23] A. Winkelmann, G. Nolze, M. Vos, F. Salvat-Pujol, W.S.M. Werner, Physics-based simulation models for EBSD: Advances and challenges, IOP Conf. Ser. Mater. Sci. Eng. 109 (2016). doi:10.1088/1757-899X/109/1/012018.

[24] P.T. Brewick, S.I. Wright, D.J. Rowenhorst, NLPAR: Non-local smoothing for enhanced EBSD pattern indexing, Ultramicroscopy. 200 (2019) 50–61. doi:10.1016/j.ultramic.2019.02.013.

[25] R. Guo, M. Ahn, H.Z. Hongtu Zhu, Spatially Weighted Principal Component Analysis for Imaging Classification, J. Comput. Graph. Stat. 24 (2015) 274–296. doi:10.1080/10618600.2014.912135.

[26] C. Eckart, G. Young, The approximation of one matrix by another of lower rank, Psychometrika. 1 (1936) 211–218. doi:10.1007/BF02288367.

[27] M.F. Møller, A scaled conjugate gradient algorithm for fast supervised learning, Neural Networks. 6 (1993) 525–533. doi:10.1016/S0893-6080(05)80056-5.

[28] R. Hielscher, F. Bartel, T.B. Britton, Gazing at crystal balls: Electron backscatter diffraction pattern analysis and cross correlation on the sphere, Ultramicroscopy. 207 (2019) 112836. doi:10.1016/j.ultramic.2019.112836.

[29] A.P. Day, Spherical EBSD, J. Microsc. 230 (2008) 472–486. doi:10.1111/j.1365-2818.2008.02011.x.

[30] A. Foden, D.M. Collins, A.J. Wilkinson, T.B. Britton, Indexing electron backscatter diffraction patterns with a refined template matching approach, Ultramicroscopy. 207 (2019) 112845.

doi:10.1016/j.ultramic.2019.112845.

[31]  T.B. Britton, J. Jiang, Y. Guo, A. Vilalta-Clemente, D. Wallis, L.N. Hansen, A. Winkelmann, A.J. Wilkinson, Tutorial: Crystal orientations and EBSD - Or which way is up?, Mater. Charact. 117 (2016) 113–126. doi:10.1016/j.matchar.2016.04.008.

[32]  A. Winkelmann, Dynamical effects of anisotropic inelastic scattering in electron backscatter diffraction, Ultramicroscopy. 108 (2008) 1546–1550. doi:10.1016/j.ultramic.2008.05.002.

[33]  M. Inokuti, S.T. Manson, Cross sections for inelastic scattering of electrons by atoms: selected topics related to electron microscopy, in: United States, 1982.

[34]  A. Winkelmann, M. Vos, The role of localized recoil in the formation of Kikuchi patterns, Ultramicroscopy. 125 (2013) 66–71. doi:10.1016/j.ultramic.2012.11.001.

[35]  B.H. Savitzky, L.A. Hughes, S.E. Zeltmann, H.G. Brown, S. Zhao, P.M. Pelz, E.S. Barnard, J. Donohue, L.R. DaCosta, T.C. Pekin, E. Kennedy, M.T. Janish, M.M. Schneider, P. Herring, C. Gopal, A. Anapolsky, P. Ercius, M. Scott, J. Ciston, A.M. Minor, C. Ophus, py4DSTEM: a software package for multimodal analysis of four-dimensional scanning transmission electron microscopy datasets, (2020) 1–32.

[36]  P.A. Midgley, A.S. Eggeman, Precession electron diffraction - A topical review, IUCrJ. 2 (2015) 126–136. doi:10.1107/S2052252514022283.

[37]  C. Ophus, Four-Dimensional Scanning Transmission Electron Microscopy (4D-STEM): From Scanning Nanodiffraction to Ptychography and Beyond, Microsc. Microanal. 25 (2019) 563–582. doi:10.1017/S1431927619000497.

[38]  T. Ben Britton, D. Goran, V.S. Tong, Space rocks and optimising scanning electron channelling contrast, Mater. Charact. 142 (2018) 422–431. doi:10.1016/j.matchar.2018.06.001.

[39]  S. Zhao, R. Zhang, T. Pekin, A.M. Minor, Probing Crystalline Defects Using an EBSD-Based Virtual Dark-Field Method, Microsc. Microanal. 25 (2019) 1992–1993. doi:10.1017/s1431927619010699.

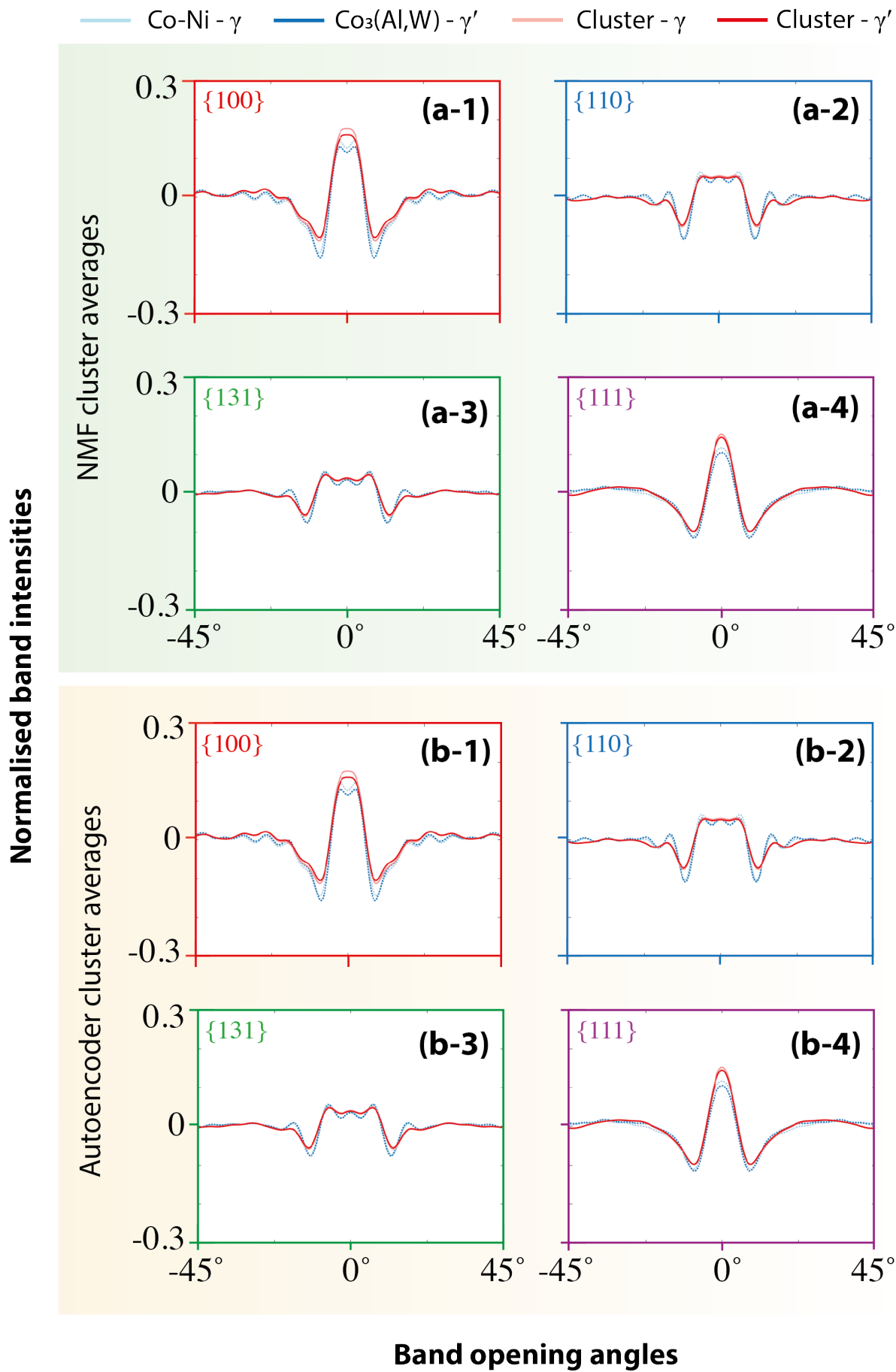

**Supplementary Figure 1**: Comparison between Kikuchi band profiles obtained from (a) averages of the NMF-identified clusters, and (b) averages of the Autoencoder-identified clusters.

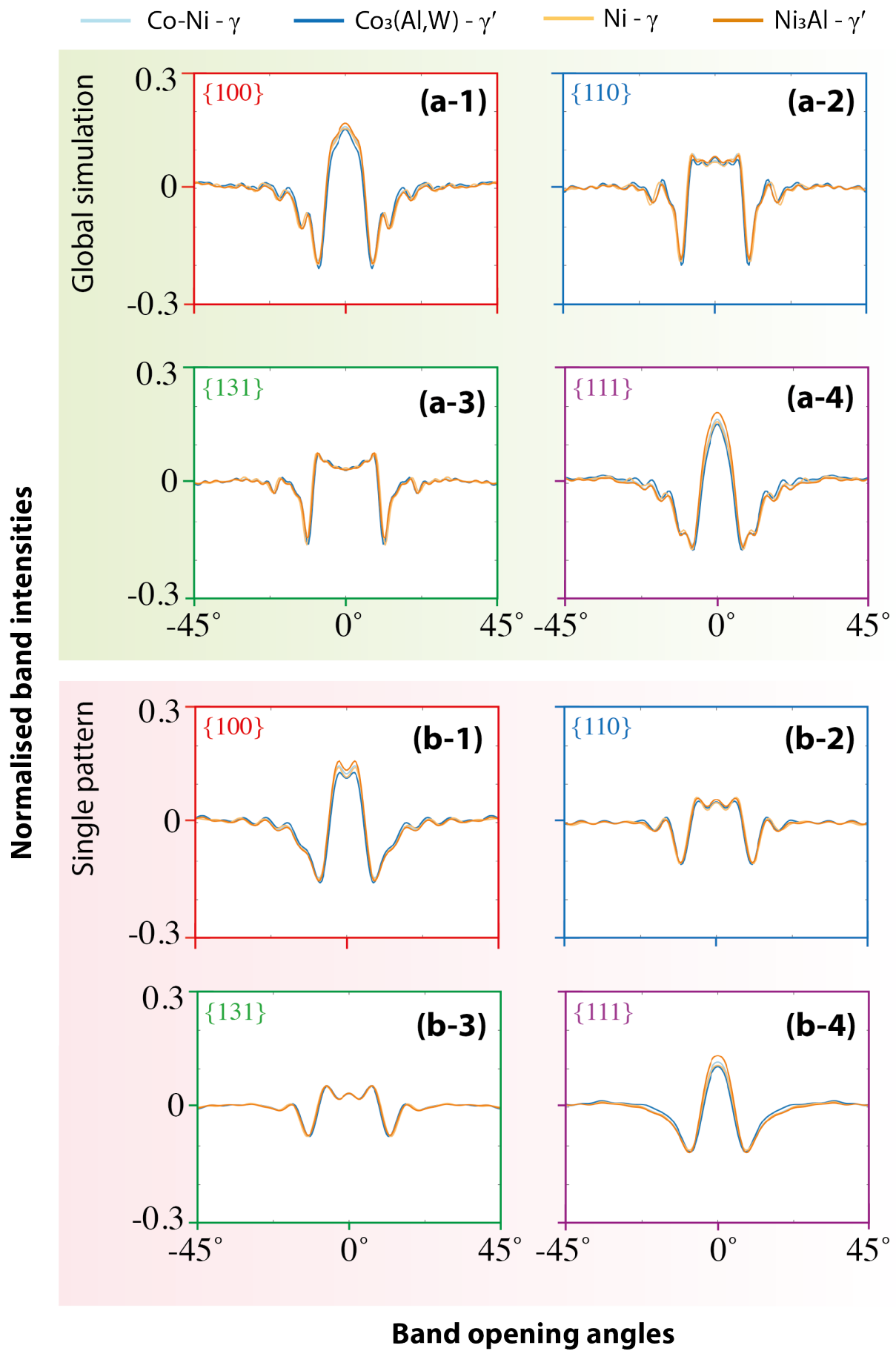

**Supplementary Figure 2**: Comparison between Kikuchi band profiles obtained from (a) the global simulation (full diffraction sphere) and (b) a re-projected single template, for both a CoNi-Co₃(Al,W) and Ni-Ni₃Al γ - γ′ simulated systems